\documentclass[12pt]{article}
\usepackage{latexsym}
\usepackage{amsmath}
\usepackage{bm}
\usepackage{graphicx,color}
\usepackage{amssymb}
\usepackage{cite}

\renewcommand{\theequation}{\thesection.\arabic{equation}}

\makeatletter
\newcommand\xleftrightarrow[2][]{%
  \ext@arrow 9999{\longleftrightarrowfill@}{#1}{#2}}
\newcommand\longleftrightarrowfill@{%
  \arrowfill@\leftarrow\relbar\rightarrow}
\makeatother

\makeatletter
\newcounter{subeqncnt}
\def\thesubeqncnt{\alph{subeqncnt}}
\def\subequations{\begingroup%
\stepcounter{equation}\edef\@tempa{\theequation}%
\let\c@equation\c@subeqncnt\c@subeqncnt\z@
\edef\theequation{\@tempa\noexpand\thesubeqncnt}}

\makeatother

\title{The Unified Soliton System as the ${\rm AdS_2}$ System} 

\author{Masahito Hayashi\thanks{masahito.hayashi@oit.ac.jp}\\
Osaka Institute of Technology, Osaka 535-8585, Japan\\
Kazuyasu Shigemoto\thanks{shigemot@tezukayama-u.ac.jp} \\
Tezukayama University, Nara 631-8501, Japan\\
Takuya Tsukioka\thanks{tsukioka@bukkyo-u.ac.jp}\\
Bukkyo University, Kyoto 603-8301, Japan\\
}
\date{\empty}

\begin{document}

\maketitle
\abstract{
We study the Riemann geometric approach to be aimed at unifying 
soliton systems. 
The general two-dimensional Einstein equation with 
a constant negative scalar curvature becomes an integrable 
differential equation. 
We show that such Einstein equation
includes KdV/mKdV/sine-Gordon equations.
}

\section{Introduction} 
\setcounter{equation}{0}

The discovery of the soliton~\cite{Gardner,Lax,Zakhrov} has given 
the breakthrough to exactly solvable non-linear equations. 
There have been many interesting 
developments to understand soliton systems  
such as the AKNS formulation~\cite{Ablowitz,Crampin,Sasaki}, 
the B\"{a}cklund
transformation~\cite{Wahlquist,Wadati1,Wadati2}, 
the Hirota equation~\cite{Hirota1,Hirota2}, 
the Sato theory~\cite{Sato},
the vertex construction of the soliton solution~\cite{Lepowsky,Date}, 
and the Schwarzian type mKdV/KdV equation~\cite{Weiss}. 

Our characterization of the soliton system is that it is a system of integrable 
non-linear differential equations, 
which have not only some exact solutions but 
also $N$-soliton solutions i.e.\ systematically obtained infinitely many solutions.
The  KdV/mKdV/sine-Gordon systems are such ones, and we 
study common structures for these soliton systems.

In our previous paper~\cite{Hayashi1,Hayashi2}, we have characterized 
such a soliton system as it has the local GL(2,$\mathbb R$)  self gauge
 symmetry, where the local gauge parameter is connected with the 
gauge potential. 
We have pointed out that 
a special local self gauge transformation becomes
 the B\"{a}cklund transformation. Combining various B\"{a}cklund 
transformations,  
we have the algebraic
relation, which becomes the addition formula to construct the $N$-soliton solution
from various 1-soliton solutions.
In our approach, the mechanism to be able to construct $N$-soliton solutions 
for the non-linear 
soliton systems comes from the GL(2,$\mathbb R$) group (=M\"{o}bius group) 
structure for such systems.

So far almost all soliton systems, which admit $N$-soliton solutions,  
are restricted only to two-dimensional models, and the Lie group
structure of such soliton systems is restricted to the rank one Lie
group GL(2,$\mathbb R$)${\rm /\mathbb{Z}_2}$  $\cong$   ${\rm SO(2,1)}$
$\cong$ ${\rm SU(1,1)/\mathbb{Z}_2}$. We will study in this paper the 
reasons of the restrictions above 
from the Riemann geometric approach.

The sine-Gordon equation is a well-known integrable system from 
old days~\cite{Bianchi}.
 In the theory of the curved surface, the fundamental equation 
 is  known as the Gauss-Weingarten equation. The integrability 
condition of this Gauss-Weingarten equation is given as the 
Gauss-Codazzi formula. The sine-Gordon equation comes from 
this Gauss-Codazzi formula  for the pseudo-sphere.
However, the approach from the curved surface is more complicated 
than that from the Riemann geometry, so that the approach 
from the curved surface seems to be 
difficult to generalize it into the higher dimensional and higher Lie group
symmetric soliton systems. Then we use the Riemann geometric
approach to the soliton systems in this paper.  

We will show in this paper, i) the general two-dimensional Einstein equation 
with  a constant negative scalar curvature, that is ${\rm AdS_2}$ system, 
becomes an integrable differential equation. ii) such Einstein
equation includes the unified soliton systems of the KdV/mKdV/sine-Gordon 
equations.
 
\section{Riemann geometric approach 
to two-dimensional soliton system as AdS$_{\bm{2}}$ system}
\setcounter{equation}{0}

We start with the general two-dimensional metric in the form
\begin{eqnarray}
{\rm d}s^2=f(x,t) {\rm d}t^2 + g(x,t) {\rm d}x^2 +2 h(x,t){\rm d}t\, {\rm d}x , 
\label{2e1}
\end{eqnarray}
where $g_{00}(x,t)=f(x,t)$, $g_{11}(x,t)=g(x,t)$, 
$g_{01}(x,t)=g_{10}(x,t)=h(x,t)$.
In two dimensions, due to the first Bianchi identity of the Riemann tensor, we 
have 
\begin{equation}
R_{ij}(x,t)=K(x,t) g_{ij}(x,t), 
\label{2e2}
\end{equation}
with
$$
K(x,t)=\frac{L(x,t)}{{\rm det} (g)}, \quad 
L(x,t)=\hat{R}_{0101}=\hat{R}_{1010}=-\hat{R}_{0110}=-\hat{R}_{1001}.
$$
We consider the case of the constant scalar curvature $R$, i.e.  
$R=2K(x,t)={\rm constant}$, and we take $K(x,t)=-1$ for 
simplicity\footnote{ 
If we choose $K(x,t)=1$, we cannot connected differential equations 
of $R=2$ with the soliton equations. In other words, 
the two-dimensional sphere with $K(x,t)=1$
has too simple structure (``No hair'') to construct the 
$N$-soliton solutions. 
}. 
In this case, the metric becomes the Einstein metric which satisfies
$R_{ij}=-g_{ij}$. 
Computing the scalar curvature with $R=-2$, we obtain 
the following differential 
equation
\begin{align}
2(f g-h^2)^2\ (R+2)
&=-2(f g-h^2)(f_{xx}+g_{tt}-2h_{xt})+f^2_{x} g+f_x g_x f-2 f_x h_x h
\nonumber\\
&\hspace*{4mm}
+g^2_{t} f+f_t g_t g-2 g_t h_t h -f_x g_t h 
+f_t g_x h 
\nonumber 
\\
&
\hspace*{4mm}
-2 f_t h_x g -2 g_x h_t f+4 h_x h_t h+4 (f g-h^2)^2 
\nonumber 
\\
&=0 .
\label{2e3}
\end{align}

Using two degrees of freedom of the 
general coordinate transformation in two dimensions, 
$t\rightarrow T(t,x)$ and
 $x\rightarrow X(t,x)$, 
two of $g_{00}$, $g_{11}$, $g_{01}$ can be set 
to be constant. Then the Einstein metric, which satisfies
$R_{ij}=-g_{ij}$, can be transformed into that of 
the pseudo-sphere $x^2+y^2-z^2=-1$~\cite{Bianchi}, which has the symmetry of SO(2,1). We parametrize it in the form
$x=\sinh{\theta} \cos{\phi}$, $y=\sinh{\theta} \sin{\phi}$,
$z=\cosh{\theta}$, which gives
\begin{align}
{\rm d}x&=\cosh{\theta} \cos{\phi}\ {\rm d}\theta
-\sinh{\theta}\sin{\theta}\ d\phi ,
\nonumber\\
{\rm d}y&=\cosh{\theta} \sin{\phi}\ {\rm d}\theta
+\sinh{\theta}\cos{\theta}\ {\rm d}\phi ,
\label{2e4}
\\
{\rm d}z&=-\sinh{\theta}\ {\rm d}\theta. 
\nonumber 
\end{align}
These lead the metric 
\begin{eqnarray}
{\rm d}s^2={\rm d}x^2+{\rm d}y^2-{\rm d}z^2
={\rm d}\theta^2+\sinh^2{\theta}\ {\rm d}\phi^2, 
\label{2e5}
\end{eqnarray}
which is one of the ${\rm AdS}_2$ parametrizations. 
In this case, $g_{00}=1$, $g_{01}=0$ are constants.
For this metric, we have 
\begin{eqnarray}
R_{00}=-1=-g_{00},\quad R_{11}=-\sinh^2{\theta}=-g_{11},\quad  
R_{01}=0, \quad R=-2.
\label{2e6}
\end{eqnarray}

In general, the Lie group symmetry of the general two-dimensional surface with 
the negative constant scalar curvature is that of the  rank one Lie group 
 GL(2,$\mathbb R$)${\rm /{\mathbb Z}_2}$ $\cong$   ${\rm SO(2,1)}$
$\cong$ ${\rm SU(1,1)/{\mathbb Z}_2}$.  
The metric Eq.(\ref{2e5})
is the real two-dimensional metric without complex structure and the 
Lie group symmetry becomes the M\"{o}bius 
group symmetry. 
If the system has the real two-dimensional metric with complex structure instead, 
the Lie group symmetry of that system becomes the infinite dimensional conformal symmetry.


\section{Integrable Condition of Surface \\
in three-dimensional Euclidean Space} 
\setcounter{equation}{0}

\subsection{Point in three-dimensional Euclidean space}

We formulate the three-dimensional geometry by the Maurer-Cartan 
formalism with the 
exterior differential form\cite{Goldberg,Flanders}. 
For any point ${\bf x}$, we attach the moving orthonormal 
basis ${\bf e_1}, {\bf e_2}, {\bf e_3} $. 
The differential 1-form ${\rm d}{\bf x}$ leads 
the structure equation 
given in
the form
\begin{align}
{\rm d}{\bf x}&=\sigma_1 {\bf e}_1+\sigma_2 {\bf e}_2+\sigma_3 {\bf e}_3
=\sigma {\bf e}, \   
\label{3e1}\\
{\rm d}{\bf e}&=\Omega \  {\bf e}, \   
\label{3e2}
\end{align}
with
$$
\sigma
=\left(\begin{array}{ccc} \sigma_1 & \sigma_2 & \sigma_3
\end{array} \right), \quad 
 {\bf e}=\left(\begin{array}{c} {\bf e}_1 \\ {\bf e}_2 \\ {\bf e}_3
\end{array} \right), \quad 
\Omega
=\left(\begin{array}{ccc} 0 & \omega_{3} & -\omega_{1} \\
-\omega_{3} & 0 & -\omega_{2}\\
\omega_{1} & \omega_{2} & 0 \\
\end{array} \right).
$$
The torsion free condition is defined by
\begin{equation}
{\rm d}\sigma=\sigma \wedge \Omega,    
\label{3e3}
\end{equation} 
and the integrability conditions of Eq.(\ref{3e3}) gives 
\begin{equation}
0={\rm d} \sigma \wedge \Omega-\sigma\wedge {\rm d}\Omega
=\sigma \wedge (\Omega \wedge \Omega -{\rm d}\Omega).  
\label{3e4}
\end{equation}
By using the first Bianchi identity, the general solution of Eq.(\ref{3e4})  is given by 
\begin{equation}
{\rm d}\Omega-\Omega \wedge \Omega=\hat{R} \sigma \wedge \sigma,  
\label{3e5}
\end{equation}
where $(\hat{R})_{ijkl}=\hat{R}_{ijkl}$ is the Riemann tensor. This is
called 
as the curvature condition.

\subsection{Point on surface in three-dimensional Euclidean space}

We formulate the unified two-dimensional soliton system as the system of the 
negative constant surface in this Maurer-Cartan geometry~\cite{Crampin,Sasaki}.
We choose the normal vector of the surface as ${\bf e_3}$.
Then the point on the surface is given by $\sigma_3=0$. Further, we have
$\omega_1=0$, $\omega_2=0$ from Eq.(\ref{3e3}).
The structure equation now becomes
\begin{equation}
{\rm d}{\bf x}=\sigma_1 {\bf e}_1+\sigma_2 {\bf e}_2, \qquad  
{\rm d}{\bf e}_1=\omega_{3} {\bf e}_2, \qquad  
{\rm d}{\bf e}_2=-\omega_{3} {\bf e}_1.
\label{3e8}
\end{equation}
%
The integrability condition turns to be 
\begin{equation}
{\rm d}\sigma_1=\omega_3 \wedge \sigma_2, \qquad 
{\rm d}\sigma_2=-\omega_3 \wedge \sigma_1, \qquad 
{\rm d}\omega_3=-K \sigma_1 \wedge \sigma_2, 
 \label{3e11}
\end{equation}
%
where we denote $K=-\hat{R}_{0101}$ as the Gauss curvature. Here we consider 
the pseudo-sphere, that is, the constant negative curvature and we take 
$K=-1$ for simplicity. Then we have the integrability condition in the form 
\begin{align}
{\rm d}\sigma_1&=\omega_3 \wedge \sigma_2,
 \label{3e12}\\
{\rm d}\sigma_2&=-\omega_3 \wedge \sigma_1,
 \label{3e13}\\
{\rm d}\omega_3&= \sigma_1 \wedge \sigma_2.
 \label{3e14}
\end{align}
This integrability condition can be expressed in the 
Sasaki's $2\times 2$ matrix form \cite{Sasaki}
\begin{equation}
{\rm d}\hat{\Omega}=\hat{\Omega} \wedge \hat{\Omega}, \quad 
\mbox{with}\quad  
\hat{\Omega}
=\frac{1}{2}\left(\begin{array}{cc} 
-\sigma_2 &  \omega_3+\sigma_1\\
 -\omega_3+\sigma_1 & \sigma_2 
\end{array} \right).
\label{3e15}
\end{equation}
%
\subsection{General integrable differential equation}

We start with the general 1-forms $\sigma_1$, $\sigma_2$,
and we construct 
$\omega_3$ from Eq.(\ref{3e12}) and Eq.(\ref{3e13}). 
Then Eq.(\ref{3e14}) gives the integrable 
differential equation.  We put
\begin{equation}
\sigma_1=f_1{\rm d}t+f_2 {\rm d}x, \quad 
\sigma_2=g_1 {\rm d}t+g_2 {\rm d}x, \quad   
\omega_3=h_1 {\rm d}t+h_2 {\rm d}x. 
\label{3e16} 
\end{equation}
From Eq.(\ref{3e12}) and Eq.(\ref{3e13}), we have
\begin{equation}
\omega_3=
\frac{(-f_{1,x}+f_{2,t}) f_1+(-g_{1,x}+g_{2,t}) g_1}
                       {f_1 g_2-f_2 g_1}\, {\rm d}t 
+\frac{(-f_{1,x}+f_{2,t}) f_2+(-g_{1,x}+g_{2,t}) g_2}
                       {f_1 g_2-f_2 g_1}\, {\rm d}x .
\label{3e17}
\end{equation}
Then Eq.(\ref{3e14}) gives the following integrable differential
equation
\begin{align}
&
\frac{\partial}{\partial x}\Big\{
\frac{(-f_{1,x}+f_{2,t}) f_1+(-g_{1,x}+g_{2,t}) g_1}
                       {f_1 g_2-f_2 g_1} \Big\} 
\nonumber\\
&
\quad 
-\frac{\partial}{\partial t}\Big\{
\frac{(-f_{1,x}+f_{2,t}) f_2+(-g_{1,x}+g_{2,t}) g_2}
                       {f_1 g_2-f_2 g_1}  \Big\} 
+f_1 g_2-f_2 g_1=0.
\label{3e18}
\end{align}
In this case, the metric is given by
\begin{align}
{\rm d}s^2&=({\rm d}{\bf x})^2=(\sigma_1 {\bf e}_1+\sigma_2 {\bf e}_2)^2
={\sigma_1}^2+{\sigma_2}^2 
\nonumber\\
&=(f_1^2+g_1^2){\rm d}t^2+(f_2^2+g_2^2){\rm d}x^2
+2(f_1 f_2+g_1 g_2){\rm d}t\, {\rm d}x \nonumber\\
&=f {\rm d}t^2+g {\rm d}x^2+2h {\rm d}t\, {\rm d}x,
\label{3e19}
\end{align}
where $f=f_1^2+g_1^2$, $g=f_2^2+g_2^2$, $h=f_1 f_2+g_1 g_2$. 
If we compare Eq.(\ref{2e3}) with Eq.(\ref{3e18}), it gives the 
same differential equation, which means that Eq.(\ref{2e3})
is the general integrable differential equation.


\section{KdV/mKdV/sine-Gordon equations as 
AdS$_{\bm 2}$ differential equation}   
\setcounter{equation}{0}
\subsection{sine-Gordon equation}

The AKNS formalism of the sine-Gordon equation
\begin{equation}
u_{xt}=\sin(u),
\label{4e1}
\end{equation}
is given in the form
\begin{equation}
{\rm d}\hat{\Omega}=\hat{\Omega} \wedge \hat{\Omega}, \quad 
\mbox{with}\quad 
\hat{\Omega}
=\left(\begin{array}{cc} 
 \eta {\rm d}x+\cos(u) {\rm d}t/4\eta &  
-u_x {\rm d}x/2+\sin(u) {\rm d}t/4 \eta\\
 u_x {\rm d}x/2+\sin(u) {\rm d}t/4\eta & -\eta {\rm d}x -\cos(u) {\rm d}t/4 \eta 
\end{array} \right).
\label{4e2} 
\end{equation}
We check how the integrability conditions Eqs.(\ref{3e12})-(\ref{3e14}) are 
satisfied. First, Eq.(\ref{3e12}) and Eq.(\ref{3e13}) are identically satisfied, 
and Eq.(\ref{3e14}) gives the sine-Gordon equation $u_{xt}=\sin(u)$. Then 
the integrability conditions of the sine-Gordon equation are in the 
standard form as Eqs.(\ref{3e16})-(\ref{3e18}). 
In this case, we have $\sigma_1=\sin(u) {\rm d}t/2$, 
$\sigma_2=-2 {\rm d}x-\cos(u){\rm d}t$ 
and the metric becomes  
\begin{align}
{\rm d}s^2&=({\rm d}{\bf x})^2=(\sigma_1 {\bf e}_1+\sigma_2 {\bf e}_2)^2
=\sigma_1^2+\sigma_2^2 
\nonumber\\
&=\frac{({\rm d}t)^2}{4}+4{\rm d}x^2+2 \cos(u) {\rm d}t\, {\rm d}x
\nonumber 
\\
&=({\rm d}\tilde{t})^2+({\rm d}\tilde{x})^2
+2 \cos(u) {\rm d}\tilde{t}\, {\rm d}\tilde{x}.
\label{4e3}
\end{align}
where $\tilde{t}=t/2$, $\tilde{x}=2x$. 
This is the standard metric 
of the sine-Gordon equation.
Calculating the scalar curvature of this metric, and taking the negative 
constant curvature, we have 
\begin{eqnarray}
K=\frac{R}{2}=-1=-\frac{u_{xt}}{\sin(u)} , 
\label{4e4}
\end{eqnarray}
which gives the sine-Gordon equation Eq.(\ref{4e1}).

\subsection{mKdV equation}

The AKNS formalism of the mKdV equation
\begin{equation}
 u_{t}+6 u^2 u_x+u_{xxx} =0, 
\label{4e5}
\end{equation}
is given by
$$
{\rm d}\hat{\Omega}=\hat{\Omega} \wedge \hat{\Omega}, 
$$
with 
\begin{equation}
\hat{\Omega}
=\left(\begin{array}{cc} 
\eta {\rm d}x-(4 \eta^3+2 \eta u^2) {\rm d}t 
& u {\rm d}x-(u_{xx}+2 \eta u_x +4 \eta^2 u +2 u^3){\rm d}t\\
-u {\rm d}x+(u_{xx}-2 \eta u_x+4 \eta^2 u +2 u^3) {\rm d}t 
& -\eta {\rm d}x+(4 \eta^3+2 \eta u^2){\rm d}t 
\end{array} \right).
\label{4e6}
\end{equation}
We check how the integrability conditions Eqs.(\ref{3e12})-(\ref{3e14}) are 
satisfied. First, Eq.(\ref{3e12}) and Eq.(\ref{3e13}) are identically satisfied, 
and Eq.(\ref{3e14}) gives the mKdV equation $u_t+6 u^2 u_x+u_{xxx}=0$.
Then the integrability conditions of the mKdV equation are in the standard form.
We take $\eta=1$ for simplicity and we have $\sigma_1=-4u_x {\rm d}t$, 
$\sigma_2=-2 {\rm d}x+4(u^2+2) {\rm d}t$ and we 
have the metric 
\begin{equation}
{\rm d}s^2=\sigma_1^2+\sigma_2^2 
=4\Big\{4 (u_x^2+(u^2+2)^2 ) {\rm d}t^2
+{\rm d}x^2-4 (u^2+2) {\rm d}t\, {\rm d}x \Big\}.
\label{4e7}
\end{equation}
Calculating the scalar curvature of this metric, and taking the negative 
constant curvature, we have 
\begin{equation}
K=\frac{R}{2}=-1=-\frac{2 u_{x}+32(u_t+6 u^2 u_x+u_{xxx})}{2 u_x} ,
\label{4e8}
\end{equation}
which gives the mKdV equation Eq.(\ref{4e5}).

\subsection{KdV equation}

For the AKNS formalism of the KdV equation
\begin{equation}
 u_{t}+6 u u_x+u_{xxx} =0  ,
\label{4e9}
\end{equation}
we use the Sasaki's form~\cite{Sasaki} to obtain the 
simple expression of the integrable geometrical differential equation. 
Then we take    
$$
{\rm d}\hat{\Omega}=\hat{\Omega} \wedge \hat{\Omega}, 
$$
with
\begin{equation}
\hat{\Omega}
=\left(\begin{array}{cc} 
 -u_x {\rm d}t & (u-\eta^2) {\rm d}x-(u_{xx}+2u^2+2\eta^2 u -4\eta^4)
  {\rm d}t\\
 -{\rm d}x+(2 u+ 4 \eta^2 ) {\rm d}t  & u_x {\rm d}t 
\end{array} \right).
\label{4e10}
\end{equation}
We check how the integrability conditions Eqs.(\ref{3e12})-(\ref{3e14}) are 
satisfied. First, Eq.(\ref{3e12}) is identically satisfied. While Eq.(\ref{3e13}) 
is not identically satisfied but gives the KdV equation $u_t+6 u u_x+u_{xxx}=0$.
Eq.(\ref{3e14}) gives the KdV equation. 
Then the integrability conditions of the KdV equation are not in the standard form.
We take $\eta=1$ for simplicity and we have  
$\sigma_1=(u-2){\rm d}x-(u_{xx}+2u^2-8){\rm d}t$, 
$\sigma_2=2 u_x {\rm d}t$ and we 
have the metric 
\begin{align}
{\rm d}s^2&=\sigma_1^2+\sigma_2^2 
\nonumber\\
&=\Big\{(u_{xx}+2u^2-8) ^2+4 u_x^2 \Big\}{\rm d}t^2
+ (u-2)^2 {\rm d}x^2-2(u-2) (u_{xx}+2 u^2-8) {\rm d}t\, {\rm d}x.
\label{4e11}
\end{align}
Denoting $P=u_t+6u u_x +u_{xxx}$, $Q=P_x=u_{xt}+6u
u_{xx}+6u_x^2+u_{xxxx}$, 
and after calculating the scalar curvature, we have
\begin{align}
&2u_x^3 (u-2)^{3} \times (R+2)
\nonumber\\
&=\Big\{ -u_{xxxx} u^2 +4 u_{xxxx} u -4 u_{xxxx}-u_{xxx} u_x u  +2 u_{xxx} u_x
+u_{xx}^2 u -2 u_{xx}^2 
\nonumber\\
&\hspace*{5mm}
 +u_{xx} u_x^2 -4 u_{xx} u^3 +20 u_{xx} u^2-32 u_{xx} u
+16 u_{xx} -6 u_x^2 u^2+24 u_x^2 u-24 u_x^2\Big\} P 
\nonumber\\
&+\Big\{
-u_{xx} u_x u +2 u_{xx} u_x -2 u_x u^3 +4 u_x u^2+8 u_x u -16 u_x\Big\}Q.
\label{4e12}
\end{align}
Then if the KdV equation Eq.(\ref{4e9}) is satisfied, we have $P=0$ and 
$Q=0$, which gives 
$R/2=K=-1$. 

However, the opposite is not always true, that is, 
even if  $R/2=K=-1$ is satisfied, we have more general differential equation 
which 
contains the KdV equation as the special case.
The reason of this property will come from the fact that
Eq.(\ref{3e12}) is not identically satisfied but gives the KdV equation itself.
In the non-linear system, it is generally difficult to have the one-to-one 
correspondence between two integrable systems.
For example, KdV system and mKdV system is connected by 
the Miura transformation $u=\pm i v_x + v^2$, 
\begin{equation}
u_t+6 u u_x +u_{xxx}=(\pm i \partial_x + 2 v)(v_t+6 v^2 v_x+v_{xxx}) , 
\label{4e13}
\end{equation}
which means that if $v$ satisfies the mKdV equation, $u$ satisfies  the KdV equation.
But the opposite is not always true. That is, if  $u$ satisfies the KdV 
equation, $v$ satisfies  
more general differential equation 
which contains the 
mKdV equation as the special case. 
In this KdV case, the integrability conditions 
Eqs.(\ref{3e12})-(\ref{3e14})  are not in the standard form. 
After the two-dimensional general coordinate transformation, 
we will be able to make the integrability conditions of the KdV equation in the standard form.

\section{Summary and discussions} 
\setcounter{equation}{0}

We have studied the Riemann geometric approach to the 
unified soliton systems, KdV/mKdV/

\noindent
sine-Gordon equations.
We have found that the general two-dimensional Einstein equation with 
a constant negative scalar curvature becomes the integrable 
differential equation.
Furthermore, we have explicitly shown that such Einstein equation includes
KdV/mKdV/sine-Gordon equations. 

We consider that a common nature of the soliton systems 
is the Lie group structure of them.
Through the Mo\"{o}bius group, many interesting approaches 
to the soliton systems are connected. The conventional geometrical 
approach to the soliton systems is to formulate the integrable 
systems in context of two-dimensional curved surface. 
In order to use 
that formalism, one must construct the complicated detailed AKNS 
operators Eq.(\ref{4e2}), Eq.(\ref{4e6}) and Eq.(\ref{4e10}) 
for each soliton systems. 
In this approach, it seems difficult
to find common features of the soliton systems, so that 
it is difficult to generalize to the higher dimensional and 
higher symmetric soliton systems.
Our geometrical approach in this paper is the Riemann geometric 
approach, which is easier to understand a common nature of the 
soliton systems in terms of the constant negative scalar curvature 
of the two-dimensional Einstein manifold, which is the simple 
realization of the 
GL(2,$\mathbb R$)${\rm /\mathbb{Z}_2}$  $\cong$   ${\rm SO(2,1)}$
$\cong$ M\"{o}bius group. 
One can find literatures in which authors  
use the Lie group symmetry to obtain the 
new solution for the differential equations such as the modified 
Klein-Gordon equation\cite{Paliathanasis1}, which is similar 
to our approach. See also the related papers\cite{Paliathanasis2}.

The reason why the two-dimensional Einstein equation with the negative
 constant scalar curvature becomes the integrable equation 
might be that such Einstein equation has the Einstein metric.
 Even in the higher dimensional and higher Lie group symmetric
soliton equations, the existence of the Einstein metric may be 
essential.
Then the Einstein equation for 
the ${\rm AdS}_n={\rm SO}(2, n-1)/{\rm SO}(1, n-1)$ 
 system and/or the Einstein equation for the K\"ahler-Einstein
 manifold, both of these systems have the Einstein metric,  
may be the candidate of the higher dimensional soliton system.
Our Riemann geometric approach to the soliton systems may give 
the non-pertubative approach
to the superstring theory through the K\"{a}hler-Einstein manifolds 
or the Calabi-Yau varieties.

The complex one-dimensional Einstein equation for the 
K\"ahler-Einstein manifold becomes the real two-dimensional
Liouville equation  
$$
\left(\frac{\partial^2}{\partial t^2}
-\frac{\partial^2}{\partial x^2}\right)u(x,t)=\exp(u(x,t)), 
$$
which is the integrable differential equation~\cite{Das}. 

\vspace{10mm}


\end{document}